\documentclass[a4paper,11pt]{article}
\usepackage{graphicx}
\usepackage{dcolumn}
\usepackage{bm}
\usepackage{graphicx,amssymb,amsmath,amsfonts,epsfig}
\usepackage[usenames,dvipsnames]{color}
\usepackage{subfigure}
\title{Horizon Areas and Logarithmic Correction to the Charged Accelerating Black Hole Entropy}
\author{Parthapratim Pradhan~\footnote{pppradhan77@gmail.com}\\ 
{\it Department of Physics}\\
{\it Hiralal Mazumdar Memorial College For Women}\\
{Dakshineswar, Kolkata-700035, India}}
\date{30-12-2018}

\begin{document}

\maketitle



\begin{abstract}

It has been shown by explicit and exact calculation that the geometric product formula i.e. horizon area (or entropy) product 
formula of outer horizon (${\cal H}^{+}$) and inner horizon (${\cal H}^{-}$) for charged accelerating black hole (BH) should 
\emph{ neither be mass-independent nor it be quantized}. This implies that the horizon area (or entropy ) product is 
mass-independent conjecture has been~\emph{broken down} for charged accelerating BH. This also further implies that the
mass-independent feature of the area product of ${\cal H}^{\pm}$ is \emph{not} a generic feature at all. We also compute 
that the \emph{Cosmic-Censorship-Inequality} for this BH. Indeed it is violated for this BH.  
Moreover, we compute the specific heat for this BH to determine 
the local thermodynamic stability of this BH. Under certain criterion, the BH shows the second order phase transition. 
Furthermore, we compute logarithmic corrections to the entropy for the said BH due to small statistical fluctuations 
around the thermal equilibrium. 
\end{abstract}


\maketitle

\section{Introduction}
Perhaps, BHs are the most facinating objects in the universe. They are the direct consequences of Einstein's 
general relativity. They could be used as a tool for testing strong gravity and beyond. The most general class 
of BHs are characterized by three parameters namely the mass, charge and spin parameter. They are described 
by Kerr-Newman family of BH in $3+1$ dimensions. It is now well established by fact that BH is a thermal 
object because it has characterized by thermodynamic variables like  temperature, entropy etc.~\cite{bk73,bcw73}. 
New thermodynamic product relations~[particularly the horizon area (or entropy) product relations] of 
event horizon~(EH)  area and Cauchy horizon ~(CH) area  for several class of BHs have been found 
universal~\cite{ah09,cgp11,finn,castro12,det12,val13,chen12,cr79}.

This relation is said to be universal because of the product of area of multihorizons particularly two 
physical horizons namely ${\cal H}^{\pm}$ is mass-independent.  When a thermodynamic product relation is 
satisfied this criterion then it is said to be a universal quantity in BH thermodynamics.
For example, in case of Kerr-Newman (KN) BH~\cite{ah09} which is an electrovacuum solution of 
Einstein's equations, it has been shown that the product of inner horizon (IH) area and outer 
horizon (OH) area should read 
\begin{eqnarray}
{\cal A}_{-} {\cal A}_{+} &=& 64 \pi^2 J^2+16 \pi^2 Q^4 ~,\label{pr1}
\end{eqnarray}
where ${\cal A}_{-}$ and ${\cal A}_{+}$ are area of CH and EH's respectively. This relation indicates 
that the universal product depends only quantized angular momentum and quantized charges 
respectively~\cite{ah09,cgp11,finn,castro12,det12,val13,chen12,cr79}. When the charge parameter 
$Q=0$, one obtains the area product 
for Kerr BH 
\begin{eqnarray}
{\cal A}_{-} {\cal A}_{+} &=& 64 \pi^2 J^2 ~,\label{pr2}
\end{eqnarray}
and it indicates that the universal product depends only quantized angular momentum parameter.
When the angular momentum parameter $J=0$, one obtains the area product formula for  spherically 
symmetric charged BH
\begin{eqnarray}
{\cal A}_{-} {\cal A}_{+} &=& 16 \pi^2 Q^4 ~,\label{pr3}
\end{eqnarray}
and it implies that the universal product depends only quantized charge parameter.
In the above three cases, it is indeed true that the horizon area~(or entropy) product 
formula is mass-independent thus it is universal in this sense. 
This is the only motivation behind this work and this is in fact an interesting topic in recent years in the 
scientific community particularly in the general relativity~(GR) community~\cite{ah09} and in the string theory 
community~\cite{cgp11,finn} [Also see references~\cite{castro12,det12,val13,chen12,cr79}] . 

It may be noted that for spherically-symmetric extremal charged BH~($M=Q$) the above Eq.~(\ref{pr3}) 
coalesces to the following equation
\begin{eqnarray}
 {\cal A}_{+}^2 &=& {\cal A}_{-}^2 =16 \pi^2 Q^4= 16 \pi^2 M^4 ~,\label{pr4}
\end{eqnarray}
Hence in this case the area's square of outer horizon or the area's square of inner horizon 
depends on the mass parameter.

Another motivation comes from Visser's work~\cite{mv13}~(see also~\cite{mv13,jh,ppgrg,pp14,plb,ijmpd}). 
Using this concept here we have tried to 
extend our analysis for \emph{charged accelerating anti-de Sitter (AdS) BH}. By explicit and exact 
calculation, we show that the area (or entropy) product formula of OH and IH for charged accelerating 
BH should \emph{not be mass-independent} and also it should  \emph{not to be quantized}. Thus, we conclude 
that the theorem of Ansorg-Hennig \cite{ah09} ``The area (or entropy) product formula is independent of mass'' 
is \emph{not} universal for charged accelerating BH. Moreover, we study other thermodynamic properties 
particularly the local thermodynamic stability by computing the specific heat. Under appropriate condition 
the BH possessess second order phase transition.   

One aspect is that the leading-order logarithmic corrections to BH entropy due to quantum fluctuations 
around the thermal equilibrium BH temperature for charged AdS BH has not been studied previously, we compute 
here the logarithmic corrections to Bekenstein-Hawking entropy for the said BH and it appears to be a generic 
feature of the BH. It should be noted that we have assumed that the BH is a thermodynamic system which is in 
equilibrium at Bekenstein-Hawking temperature.

The another strong motivation comes from the work of Cveti\v{c} et al. \cite{cgp11}, where the authors suggested that 
if any how the \emph{cosmological parameter is quantized} then the area (or entropy) product relations for 
rotating BH in $D=4$ and $D>4$ should provide a ``looking glass for probing the microscopics of general BHs''. 
Thus it is quite interesting to investigate the area (or entropy) product formula after incorporating the cosmological 
constant.

The interesting property of the charged AdS BH is that it has an accelerating horizon and the OH posessess conical 
singularity \cite{ruth}.  In spite of the non-asymptotic structure, the first law of BH thermodynamics and Smarr formula 
are satisfied  for this accelerating spacetime~\cite{ruth} (See also ~\cite{asto}). But the cosmological horizon don't have
accelerating horizon. This type of BH is said to be \emph{slowly} accelerating BH. The other novel properties of this accelerating BH is that 
it is described by the $C$ metric \cite{walker,pleban,dias,griff}. Again the $C$ metric has some peculiar features in the sense that 
it accelerates by pulling with a `cosmic string' which described by a `conical deficit in the spacetime' which connects the OH 
of the BH to infinity \cite{ruth}. 

Actually the idea of vacuum $C$ metric was first given by Levi-Civita in 1918 \cite{levi}. 
Then it was rediscovered by Newman and Tamburino in 1961 \cite{tamb}, also by Robinson and Trautman \cite{rt} in same years, and 
by Ehlers and Kundt \cite{kundt} in 1963 but there were no explanation has given. First Robinson and Trautman discovered the 
interesting features of this metric that is it emits gravitational radiation. Later Podolsky et al. \cite{podo1} studied the 
gravitational and electromagnetic radiation emitted by the uniformly accelerated charged BH in AdS spacetime.

In the next section, we have given the basic characteristics of the charged accelerating BH and we have also 
derived the area functional relation in terms of  BH mass, charge, acceleration, cosmological constant. We have also 
derived the cosmic censorship inequality for this BH and finally we have computed the specific heat which determines 
the local thermodynamic stability. In the third section, we have discussed the logarithmic correction to BH entropy for 
this class of BHs. Finally in the last section,  we have given the conclusion. 

\section{Thermodynamic properties of Charged Accelerating BH:}
The metric of the charged accelerating BH \cite{walker,pleban,griff,ruth} is described by 
\begin{eqnarray}
ds^2 &= & \frac{1}{\Omega^2} \left[-{\cal F}(r) dt^2 + \frac{dr^2}{{\cal F}(r)} 
+r^2 \left(\frac{d\theta^2}{{\cal G}(\theta)} +{\cal G}(\theta) sin^2\theta \frac{d\phi^2}{K^2}\right)\right] .~\label{ac4}
\end{eqnarray}
with the gauge potential $F=dB$ and $B=-\frac{q}{r}dt$, and where 
\begin{eqnarray}
{\cal F}(r) &=& (1-\chi^2r^2)\left(1-\frac{2m}{r}+\frac{q^2}{r^2} \right)+\frac{r^2}{\ell^2}, \\
{\cal G}(\theta) &=& 1+2m \chi cos\theta+q^2\chi^2cos^2\theta .~\label{ac5}
\end{eqnarray}
and the conformal factor is $\Omega=1+\chi r cos\theta$. It determines the conformal infinity of the AdS BH. The quantities 
$m$ and $q$ are BH mass and  BH charge respectively, $\chi>0$ determines the acceleration of the BH and 
$-\frac{\Lambda}{3}=\frac{1}{\ell^2}$ where $\ell$ is the radius of the AdS BH. It should be noted that when $\chi<\frac{1}{\ell}$, 
a single BH is present with single horizon \cite{podo} whereas when $\chi>\frac{1}{\ell}$, two BHs of opposite charge are separated 
by accelerating horizon \cite{dias,krtous} and when $\chi=\frac{1}{\ell}$ is a special case and it has been explicitly described 
in \cite{emparan}. To obtain the angular coordinates as usual form on ${\cal S}^2$ we have set the restriction $m\chi<\frac{1}{2}$. 

Now we discuss the angular part of the metric and the properties of ${\cal G}(\theta)$ at the north pole $(\theta=0)$ and 
south pole $(\theta=\pi)$. In general, $K={\cal G}(\theta)$ but at north pole fixed with $K=K_{N}=1+2m\chi+q^2\chi^2$ and 
at south pole $K_{S}=1-2m\chi+q^2\chi^2$ that indicates the choices at the north pole and at the south pole are mutually 
incompatible. Thus the metric cannot be made regular at both the poles at the same time. Therefore the convention is to make 
a choice that makes it regular at one of the poles-- this leads to  a 
`conical deficit' at the other pole which is $\delta=\frac{8\pi m \chi}{1+2m\chi+q^2\chi^2}$ and which corresponds 
to a `cosmic string' \cite{ruth} with tension $\mu=\frac{\delta}{8\pi}=\frac{m \chi}{1+2m\chi+q^2\chi^2}$.

Thus the $C$ metric is described by the five physical parameters: the mass $m$, the charge $q$, the negative cosmological 
constant $-\Lambda=\frac{3}{\ell^2}$, the acceleration $\chi$ and the `tension of the cosmic string' on each axis which 
is represented by the periodicity of the angular coordinate. More discussion regarding the thermodynamic properties 
( particularly first law of BH thermodynamics \cite{asto}, Smarr mass formula, thermodynamic volume, Gibbs free energy 
and Reverse isoperimetric inequality) of the $C$ metric could be found in \cite{ruth}. 


Now we evaluate the radii of BH horizons by imposing the condition  ${\cal F}(r_{i})=0$ i.e. 
\begin{eqnarray}
\left(1-\chi^2\ell^2\right)r_{i}^4+2m\ell^2\chi^2r_{i}^3+\ell^2\left(1-\chi^2q^2\right)r_{i}^2-2m\ell^2r_{i}+q^2\ell^2 &=& 0 
~.\label{ac6}
\end{eqnarray}
Apply the Vieta's theorem, we find
\begin{eqnarray}
\sum_{i=1}^{4} r_{i} &=& -\frac{2m\ell^2\chi^2}{\left(1-\chi^2\ell^2\right)} ~.\label{eq1}\\
\sum_{1\leq i<j\leq 4} r_{i}r_{j} &=& \ell^2 \frac{\left(1-\chi^2q^2\right)}{\left(1-\chi^2\ell^2\right)}  ~.\label{eq2}\\
\sum_{1\leq i<j<k\leq 4} r_{i}r_{j} r_{k} &=& \frac{2m\ell^2}{\left(1-\chi^2\ell^2\right)} ~.\label{eq3}\\
\sum_{1\leq i<j<k<l\leq 4} r_{i}r_{j} r_{k}r_{l} &=& \frac{q^2\ell^2}{\left(1-\chi^2\ell^2\right)}   ~.\label{eq4}
\end{eqnarray}
Eliminating the mass parameter, one obtains a single mass-independent relation as 
$$
r_{1}r_{2}+\frac{q^2\ell^2}{\left(1-\chi^2\ell^2\right)r_{1}r_{2}}+\left(r_{1}+r_{2}\right)^2 \times
$$
\begin{eqnarray}
\left[r_{1}r_{2} \frac{\chi^2}{\left(1+\chi^2 r_{1}r_{2}\right)}-\frac{q^2\ell^2\chi^2}{r_{1}r_{2}\left(1-\chi^2\ell^2\right)
\left(1+\chi^2 r_{1}r_{2}\right)}-1\right]  &=& \ell^2 \frac{\left(1-\chi^2q^2\right)}{\left(1-\chi^2\ell^2\right)}  ~.\label{eq5}
\end{eqnarray}
In terms of two BH physical horizons area ${\cal A}_{i}=\frac{4\pi r_{i}^2}{K\left(1-\chi^2r_{i}^2\right)}$ 
(where $i=1$ for EH and $i=2$ for CH), the mass independent functional relationship is given by 
$$
\sqrt{\frac{K{\cal A}_{1}}{4\pi+K\chi^2 {\cal A}_{1}}} \sqrt{\frac{K{\cal A}_{2}}{4\pi+K\chi^2 {\cal A}_{2}}}+
\frac{q^2\ell^2}{\left(1-\chi^2\ell^2\right)} \sqrt{\frac{4\pi+K\chi^2 {\cal A}_{1}}{K{\cal A}_{1}}}
\sqrt{\frac{4\pi+K\chi^2 {\cal A}_{2}}{K{\cal A}_{2}}}+
$$
$$
\left(\sqrt{\frac{K{\cal A}_{1}}{4\pi+K\chi^2 {\cal A}_{1}}}+ \sqrt{\frac{K{\cal A}_{2}}{4\pi+K\chi^2 {\cal A}_{2}}}\right)^2 \times
$$
$$
\left[\sqrt{\frac{K{\cal A}_{1}}{4\pi+K\chi^2 {\cal A}_{1}}} \sqrt{\frac{K{\cal A}_{2}}{4\pi+K\chi^2 {\cal A}_{2}}}
\frac{\chi^2}{\left(1+\chi^2 \sqrt{\frac{4\pi+K\chi^2 {\cal A}_{1}}{K{\cal A}_{1}}}
\sqrt{\frac{4\pi+K\chi^2 {\cal A}_{2}}{K{\cal A}_{2}}} \right)} -1 \right] -
$$
$$
\left(\sqrt{\frac{K{\cal A}_{1}}{4\pi+K\chi^2 {\cal A}_{1}}}+ \sqrt{\frac{K{\cal A}_{2}}{4\pi+K\chi^2 {\cal A}_{2}}}\right)^2 
\times
$$
$$
\sqrt{\frac{4\pi+K\chi^2 {\cal A}_{1}}{K{\cal A}_{1}}}\sqrt{\frac{4\pi+K\chi^2 {\cal A}_{2}}{K{\cal A}_{2}}}\times
$$
\begin{eqnarray}
\left[\frac{q^2\ell^2\chi^2}{\left(1-\chi^2\ell^2\right)\left(1+\chi^2 \sqrt{\frac{4\pi+K\chi^2 {\cal A}_{1}}{K{\cal A}_{1}}}
\sqrt{\frac{4\pi+K\chi^2 {\cal A}_{2}}{K{\cal A}_{2}}}\right)}\right] &=& 
\ell^2 \frac{\left(1-\chi^2q^2\right)}{\left(1-\chi^2\ell^2\right)} \nonumber\\ 
~.\label{eq6}
\end{eqnarray}
where,
\begin{eqnarray}
K &=& 1+2m \chi +q^2\chi^2 .~\label{eq7}
\end{eqnarray}
Eq. (\ref{eq5}) is indeed mass independent but the difficulties arise when we write the expression in terms of area of the BH 
physical horizons [Eq. (\ref{eq6})] because there is a factor $K$ where a mass term $m$ is present. Therefore it is indeed true 
that the \emph{mass-independent relation has been violated} for charged accelerating BH. 
Thus the ``Ansorg-Hennig \cite{ah09} area theorem'' conjecture breaks down for charged accelerating BH. This is an another example 
we have provided in the literature that the area product of OH and IH is \emph{not} always universal.

The BH entropy~\cite{ruth} is given by 
\begin{eqnarray}
S_{i} &=& \frac{{\cal A}_{i}}{4}= \frac{\pi r_{i}^2}{K\left(1-\chi^2r_{i}^2\right)} .~\label{eq9}
\end{eqnarray}
and the electric potential on the horizon should read 
\begin{eqnarray}
{\Phi}_{i} &=& \frac{q}{r_{i}} .~\label{eq10}
\end{eqnarray}
Now the BH temperature is given by
\begin{eqnarray}
T_{i} &=& \frac{{\cal F}'(r_{i})}{4\pi}= 
\frac{1}{4\pi} \left[ \frac{2m}{r_{i}^2}-\frac{2q^2}{r_{i}^3}+2m\chi^2-2\chi^2r_{i}+\frac{2r_{i}}{\ell^2} \right]
~. \label{eq11}
\end{eqnarray}
The other useful thermodynamic relations are  
\begin{eqnarray}
T_{i} S_{i} &=& \frac{M \left(1+\chi^2r_{i}^2\right)}{2\left(1-\chi^2r_{i}^2\right)}-\frac{\Phi Q}{2\left(1-\chi^2r_{i}^2\right)}
-\frac{\chi^2r_{i}^3}{2K\left(1-\chi^2r_{i}^2\right)}+P\frac{4\pi}{3K}\frac{r_{i}^3}{\left(1-\chi^2r_{i}^2\right)}.~\label{eqc4}
\end{eqnarray}
where we have set the parameter $m=MK$, $q=QK$ and $P=-\frac{\Lambda}{8\pi}=\frac{3}{8\pi \ell^2}$. 

The Smarr-Gibbs-Duhem relation becomes
\begin{eqnarray}
M &=& 2 \left( \frac{1-\chi^2r_{i}^2}{1+\chi^2r_{i}^2}\right)\left(T_{i}S_{i}-PV\right)+
\frac{\Phi Q}{\left(1+\chi^2r_{i}^2\right)} +\frac{\chi^2r_{i}^3}{K\left(1+\chi^2r_{i}^2\right)}.~\label{ceq5}
\end{eqnarray}
The thermodynamic volume is derived to be 
\begin{eqnarray}
V &=& \left(\frac{\partial M}{\partial P}\right)_{S,Q}=\frac{4\pi}{3K}\frac{r_{i}^3}{\left(1-\chi^2r_{i}^2\right)}  
~.\label{ceq6}
\end{eqnarray} 
beacuse the mass parameter becomes
\begin{eqnarray}
M &=& \frac{1}{2K} \left[r_{i}+\frac{K^2Q^2}{r_{i}}+\frac{8\pi P}{3}\frac{r_{i}^3}{\left(1-\chi^2r_{i}^2\right)} \right]
.~\label{eqc7}
\end{eqnarray}
When the acceleration $\chi$ vanishes, one obtains indeed the result of charged AdS BH. When we have taken into the 
concept of extended phase space then the first law has taken to be the form as  
\begin{eqnarray}
 dM &=&  T_{i} dS_{i} +V dP +\Phi_{i} dQ ~. \label{fl}
\end{eqnarray}
where the  thermodynamic volume is defined in Eq.(\ref{ceq6}). 

The \emph{reverse isoperimetric inequality} indeed \emph{violated} for this BH as 
\begin{eqnarray}
 {\cal R} &=& \left(1-\chi^2 r_{+}^2 \right)^\frac{1}{6} \leq 1 ~.\label{eqc8}
\end{eqnarray}
Finally the  Gibb's free energy is defined to be 
\begin{eqnarray}
 G_{i} &=& M-T_{i}S_{i} ~. \label{eqc9}
\end{eqnarray}
where $M$ and $T_{i}S_{i}$ are defined in Eq.~(\ref{eqc7}) and Eq.~(\ref{eqc4}). After substituting 
these values one obtains
\begin{eqnarray}
 G_{i} &=& \frac{ \left(1-3\chi^2r_{i}^2\right)}{\left(1+\chi^2r_{i}^2\right)}T 
 \frac{\pi r_{i}^2}{K\left(1-\chi^2r_{i}^2\right)}
 +\frac{\Phi Q}{\left(1+\chi^2r_{i}^2\right)}
+\frac{\chi^2r_{i}^3}{K\left(1+\chi^2r_{i}^2\right)}-P\frac{8\pi}{3K}\frac{r_{i}^3}{\left(1+\chi^2r_{i}^2\right)} 
~. \label{eqc10}
\end{eqnarray}
Solving the equation $G_{i}=0$, one finds the critical temperature 
\begin{eqnarray}
T_{c} &=& \frac{P\frac{8\pi}{3K}\frac{r_{i}^3}{\left(1+\chi^2r_{i}^2\right)}
-\frac{\chi^2r_{i}^3}{K\left(1+\chi^2r_{i}^2\right)}-\frac{\Phi Q}{\left(1+\chi^2r_{i}^2\right)}}
{\frac{\pi r_{i}^2}{K\left(1-\chi^2r_{i}^2\right)}\frac{\left(1-3\chi^2r_{i}^2\right)}{\left(1+\chi^2r_{i}^2\right)} } ~\label{gf1}
\end{eqnarray}
For $T>T_{c}$, the BH should be stable  while the Gibb's free energy is minimum and for $T<T_{c}$, $G_{i}>0$.

We know the  famous~\emph{Cosmic-Censorship-Inequality} [which requires cosmic-censorship hypothesis 
~\cite{rp}~(See \cite{bray,bray1,jang,rg,gibb99})] for Schwarzschild BH is given by
\begin{eqnarray}
m  &\geq&  \sqrt{\frac{{\cal A}}{16\pi}} ~.\label{acpi}
\end{eqnarray}
It is a very challenging topic in mathematical relativity since 1973.

The above inequality~\footnote{The Penrose inequality is violated for charged accelerating BH because 
in the right side of the Eq.~(\ref{eq6}) the factor $K$ depends on the mass parameter.} for accelerating BH becomes 
\begin{eqnarray}
m  & \geq& \frac{1}{2}\sqrt{\frac{K{\cal A}_{i}}{4\pi+K\chi^2 {\cal A}_{i}}} 
 \left[ 1+q^2 \left(\frac{4\pi+K\chi^2 {\cal A}_{i}}{K{\cal A}_{i}}\right)+\frac{K{\cal A}_{i}}{4\pi \ell^2}\right]~.\label{eq12}
\end{eqnarray}
This idea had first given in 1973 by Penrose~\cite{rp}  which is an important topic in GR
which relates  ADM mass (i.e. total mass of the spacetime) and the area of the even horizon. It is called 
Cosmic-Censorship-Inequality or Cosmic-Censorship-Bound~\cite{gibb05}.
This inequality implies that it provides the lower bound on the mass (or energy) for any time-symmetric initial data
which  satisfied the famous Einstein equations with negative cosmological constant, and which also satisfied the 
\emph{dominant energy condition} which possesses no naked singularities. 

Local thermodynamic stability and  phase transitions~\cite{hp83,david12} particularly second order phase transition are 
important phenomena in BH thermodynamics and it can be determined by computing the specific heat which is calculated to be 
for accelerating BH:
\begin{eqnarray}
C_{i}  &=&2\pi r_{i}^2 \frac{\left[1-\frac{q^2}{r_{i}^2}+\frac{r_{i}^2}{\ell^2}\frac{(3-\chi^2r_{i}^2)}{(1-\chi^2r_{i}^2)^2}
\right]}{\left[\frac{4\chi^2r_{i}^4}{\ell^2(1-\chi^2r_{i}^2)^2}+\frac{r_{i}^2}{\ell^2}\left(\frac{3-\chi^2r_{i}^2}{1-\chi^2r_{i}^2}\right)
+\frac{3q^2}{r_{i}^2}-\left(1+q^2\chi^2+\chi^2r_{i}^2\right)\right]}  ~.\label{eq13}
\end{eqnarray}
Now we analyze the above expression of specific heat for different parameter space.

\emph{Case I:}

When 
\begin{eqnarray}
1+\frac{r_{i}^2}{\ell^2}\frac{(3-\chi^2r_{i}^2)}{(1-\chi^2r_{i}^2)^2} >\frac{q^2}{r_{i}^2}  \,\, \mbox{and} \nonumber\\
\frac{4\chi^2r_{i}^4}{\ell^2(1-\chi^2r_{i}^2)^2}+\frac{r_{i}^2}{\ell^2}\left(\frac{3-\chi^2r_{i}^2}{1-\chi^2r_{i}^2}\right)
+\frac{3q^2}{r_{i}^2}>\left(1+q^2\chi^2+\chi^2r_{i}^2\right)
\end{eqnarray}
the specific heat is positive  i. e. $C_{i}>0$, which indicates that  the BH is thermodynamically stable. This inequality 
is plotted in Fig. 1(a). Along the abcissa we have taken the value of horizon radius $r_{+}$ and along the ordinate 
we have taken the value of $q$.

\emph{Case II:}
When 
\begin{eqnarray}
1+\frac{r_{i}^2}{\ell^2}\frac{(3-\chi^2r_{i}^2)}{(1-\chi^2r_{i}^2)^2} >\frac{q^2}{r_{i}^2}  \,\, \mbox{and} \nonumber\\
\frac{4\chi^2r_{i}^4}{\ell^2(1-\chi^2r_{i}^2)^2}+\frac{r_{i}^2}{\ell^2}\left(\frac{3-\chi^2r_{i}^2}{1-\chi^2r_{i}^2}\right)
+\frac{3q^2}{r_{i}^2}<\left(1+q^2\chi^2+\chi^2r_{i}^2\right)
\mbox{or} \nonumber\\
1+\frac{r_{i}^2}{\ell^2}\frac{(3-\chi^2r_{i}^2)}{(1-\chi^2r_{i}^2)^2} <\frac{q^2}{r_{i}^2}  \,\, \mbox{and} \nonumber\\
\frac{4\chi^2r_{i}^4}{\ell^2(1-\chi^2r_{i}^2)^2}+\frac{r_{i}^2}{\ell^2}\left(\frac{3-\chi^2r_{i}^2}{1-\chi^2r_{i}^2}\right)
+\frac{3q^2}{r_{i}^2}>\left(1+q^2\chi^2+\chi^2r_{i}^2\right)
\end{eqnarray}
the specific heat is negative i. e. $C_{i}<0$, which implies that the BH is thermodynamically unstable. 
These two inequalities are shown in Fig. 1(b) and Fig. 1(c) .

\emph{Case III:}
When
\begin{eqnarray}
\frac{\left[\frac{4\chi^2r_{i}^4}{\ell^2(1-\chi^2r_{i}^2)^2}+\frac{r_{i}^2}{\ell^2}\left(\frac{3-\chi^2r_{i}^2}{1-\chi^2r_{i}^2}\right)
+\frac{3q^2}{r_{i}^2}\right]}{\left(1+q^2\chi^2+\chi^2r_{i}^2\right)}  &=& 1  ~.\label{eq14}
\end{eqnarray}
the specific heat $C_{i}$ diverges. It signals a second order phase transition for such BHs. It could be observed from 
the Fig.~\ref{as2}.
\begin{figure}[h]
\begin{center}
\subfigure[]{
\includegraphics[width=2.1in,angle=0]{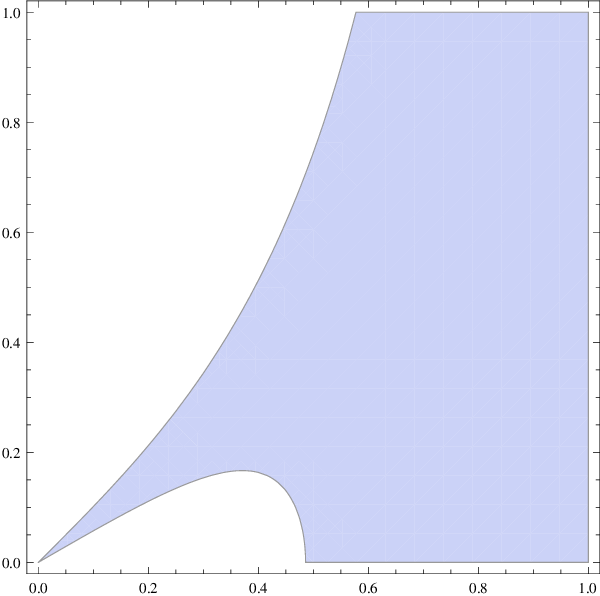}} 
\subfigure[]{
 \includegraphics[width=2.1in,angle=0]{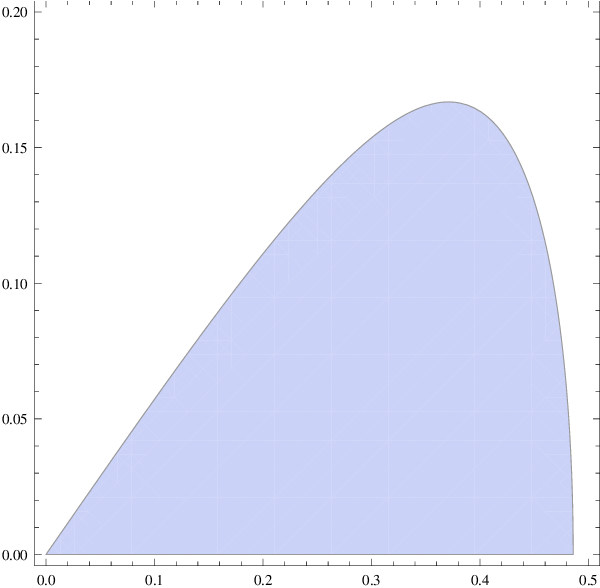}}
 \subfigure[]{
 \includegraphics[width=2.1in,angle=0]{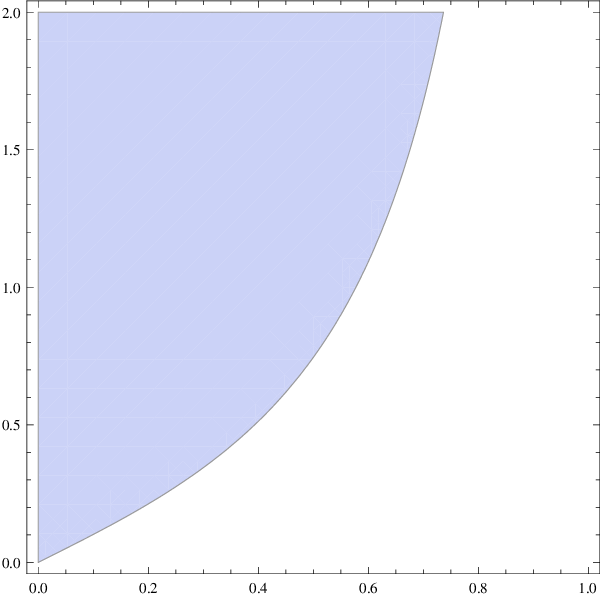}}
  \caption{\label{as1}\textit{The inequality for specific heat of the Case-I is plotted in Fig.~1(a), 
  Case-II is plotted in Fig.~1(b) and Fig.~1(c)}. Along the abcissa~($X$) we have chosen the value of 
  horizon radius $r_{+}$ and along the ordinate~($Y$) we have chosen the value of $q$. 
  We have set $\ell=\chi=1$.}
\end{center}
\end{figure}
In Fig.~\ref{as2}, we have plotted  specific heat with the horizon radius. 
From the figure one can observed that the phase transition occurs at the  
positive value of the EH radius. 
\begin{figure}[h]
 \begin{center}
 \subfigure[ ]{
 \includegraphics[width=2.1in,angle=0]{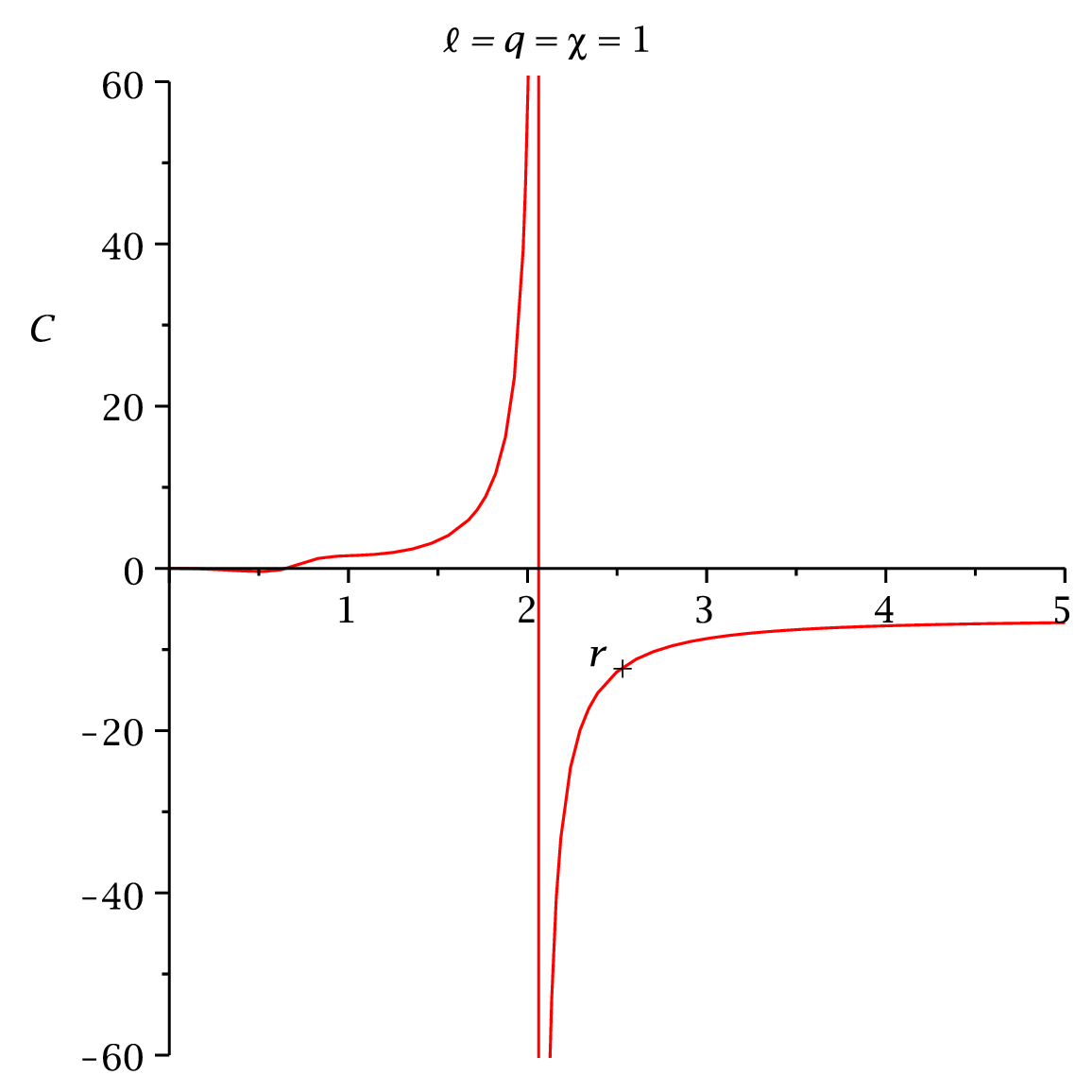}}
 \subfigure[ ]{
 \includegraphics[width=2.1in,angle=0]{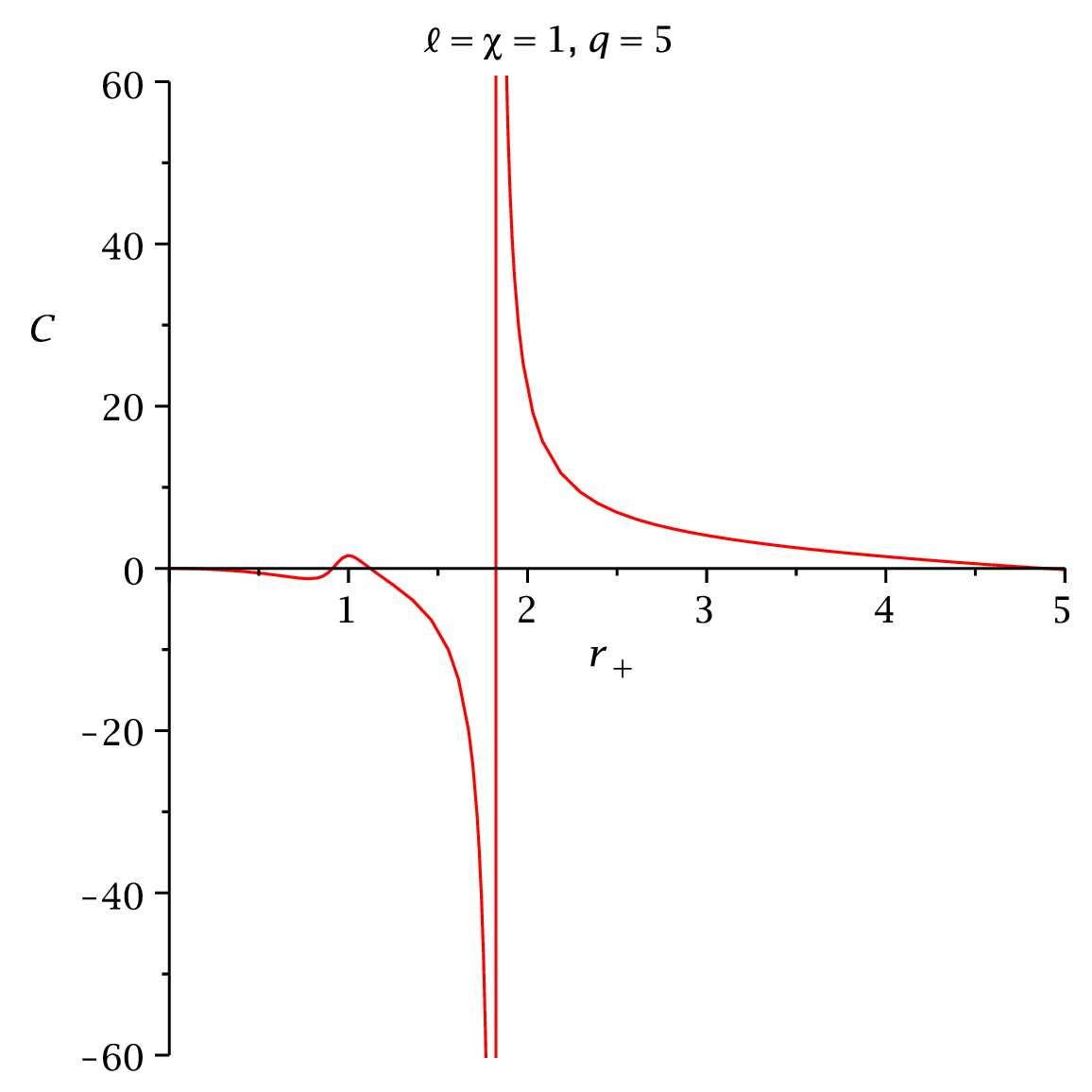}}
 \caption{\label{as2}\textit{In this figure, we have plotted the variation of specific heat~$C_{i}$
 with horizon radius~($r_{+}$) for the values $\chi=1$ and $\ell=1$. }}
\end{center}
\end{figure}

\section{Logarithmic Corrections to Entropy for charged accelerating BH:}
In this section, we should derive the logarithmic corrections to BH entropy for charged accelerating BH by
 assuming that the BH considered as a thermodynamic system which is in equilibrium at Bekenstein-Hawking 
temperature, and  due to the effects of statistical thermal fluctuations around the equilibrium. 

To derive this correction,  we should follow the classical  work of Das et al. \cite{psm}. Where the authors 
first studied the general logarithmic corrections to BH entropy for higher dimensional AdS spacetime and BTZ BH.
Since the BHs have been considered as macroscopic object compared to the Planck scale length  and this indicates 
that the logarithmic terms are much smaller compared to the Bekenstein-Hawking terms and therefore it should 
be treated as corrections.

Now we can define the canonical partition function \cite{haw77} as
\begin{eqnarray}
 Z_{i}(\beta_{i}) &=& \int_{0}^{\infty} \rho_{i}(E) e^{-\beta_{i} E} dE  ~. \label{vpa}
\end{eqnarray}
where $T_{i}=\frac{1}{\beta_{i}}$ is the temperature of ${\cal H}^{i}$. We have chosen the value of 
Boltzman constant $k_{B}$ to be unity.

The density of states can be defined as an inverse Laplace transformation of
the partition function:
\begin{eqnarray}
 \rho_{i}(E) &=& \frac{1}{2 \pi i}\int_{a-i\infty}^{a+i\infty} Z_{i}(\beta_{i})
 e^{\beta_{i} E} d\beta_{i} \\
 &=& \frac{1}{2 \pi i}\int_{a-i\infty}^{a+i\infty}
 e^{S_{i}(\beta_{i})} d{\beta_{i}}~. \label{via}
\end{eqnarray}
where $a$ is a real positive constant and 
\begin{eqnarray}
S_{i} &=&  \ln Z_{i} +\beta_{i} E  ~. \label{va}
\end{eqnarray}
is the entropy of the system near its equilibrium.

Near equilibrium of the inverse Hawking temperature $\beta_{i}=\beta_{0, i}$, we can expand the 
entropy function as
\begin{eqnarray}
S_{i}(\beta_{i}) &=&  S_{0, i}+\frac{1}{2} (\beta_{i}-\beta_{0, i})^2 S_{0, i}'' + ...  ~.\label{sba}
\end{eqnarray}
where, $S_{0, i}: =S_{i}(\beta_{0, i})$ and $S_{0, i}''=\frac{\partial^2 S_{i}}{\partial \beta_{i}^2}$
at $\beta_{i}=\beta_{0, i}$.

Putting the  Eq. (\ref{sba}) in Eq. (\ref{vpa}), one obtains
\begin{eqnarray}
\rho_{i}(E) &=& \frac{e^{S_{0, i}}}{2 \pi i}\int_{a-i\infty}^{a+i\infty}
e^\frac{\left(\beta_{i}-\beta_{0, i}\right)^2 S_{0, i}''}{2} d{\beta_{i}}~.\label{rha}
\end{eqnarray}
Let us choose $\beta_{i}-\beta_{0, i} =i y_{i}$ and setting $a=\beta_{0, i}$, $y_{i}$ is a real variable 
and evaluating a contour integral we have
\begin{eqnarray}
 \rho_{i}(E) &=& \frac{e^{S_{0, i}}}{\sqrt{2 \pi S_{0, i}''}}~.\label{psa}
\end{eqnarray}
The logarithm of $\rho_{i}(E)$ gives the corrected entropy of the thermodynamic system:
\begin{eqnarray}
{\cal S}_{i}:  &=&  \ln \rho_{i} ={\cal S}_{0, i}-\frac{1}{2} \ln S_{0, i}''+ ... ~. \label{sra}
\end{eqnarray}

Next our task is to compute $S_{0, i}''$, for this we must choose any specific form of function ${\cal S}_{i}(\beta_{i})$ 
which is modular invariant partition function  and which is also admitted an extremum at some specific value 
$\beta_{i, 0}$ of  $\beta_{i}$ followed by underlying conformal field theory (CFT) \cite{carlip1,psm}. Therefore 
the exact entropy function followed by CFT is of the form:
\begin{eqnarray}
{\cal S}_{i}(\beta_{i})=c\beta_{i}+\frac{d}{\beta_{i}}
\end{eqnarray}
where $c,d$ are constants. It can be rewritten as more general form which admits saddle point as 
\begin{eqnarray}
{\cal S}_{i}(\beta_{i})=c\beta_{i}^m+\frac{d}{\beta_{i}^n}
\end{eqnarray}
where $m, n, c, d>0$. The special case we have considered here  when $m=n=1$ and it is due to the CFT.
After some algebraic computation (See for more details \cite{psm,pp}) one can find the value of 
$S_{0, i}''=T_{i}^2 S_{0, i}$,  then we get the leading order corrections to BH entropy as 
\begin{eqnarray}
{\cal S}_{i}  &=&  \ln \rho_{i} ={\cal S}_{0, i}-\frac{1}{2} \ln \left| T_{i}^2 S_{0, i} \right|+...  ~ \label{vea}
\end{eqnarray}

Putting the values of ${\cal S}_{0, i}=\frac{\pi r_{i}^2}{K\left(1-\chi^2r_{i}^2\right)}$ and $T_{i}$, 
one can obtain the logarithmic correction to BH entropy for charged accelerating BH:
$$
{\cal S}_{i}  = \frac{\pi r_{i}^2}{K\left(1-\chi^2r_{i}^2\right)}-
\ln \left|\frac{2m}{r_{i}}-\frac{2q^2}{r_{i}^2}+2m\chi^2r_{i}-2\chi^2r_{i}^2+\frac{2r_{i}^2}{\ell^2}\right|+
$$
\begin{eqnarray}
\frac{1}{2} \ln 
\left|K\left(1-\chi^2r_{i}^2\right)\right|+...  ~ \label{eca}
\end{eqnarray}
This corrected entropy formula indicates that it is a function of horizon radius, mass parameter, charge parameter, acceleration and 
cosmological constant.

It should be noted that the product 
\begin{eqnarray}
\prod_{i=1}^{2} {\cal S}_{i}  ~\label{eqa}
\end{eqnarray}
is explicitly depends on the value of mass parameter and etc. Thus the logarithmic corrected entropy formula 
is not mass-independent and it does not quantized.

\begin{figure}[h]
 \begin{center}
 \subfigure[ ]{
 \includegraphics[width=2.1in,angle=0]{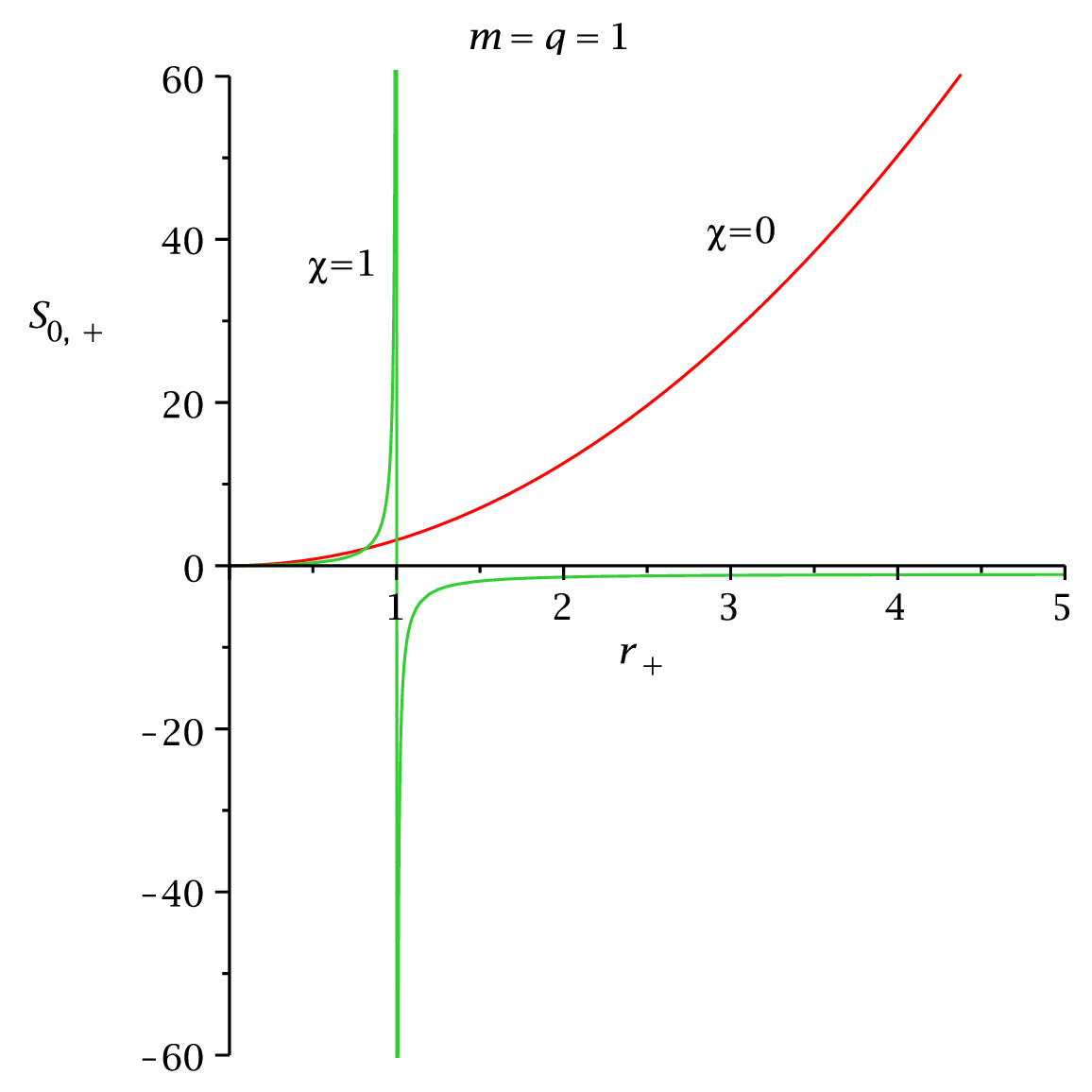}}
 \subfigure[ ]{
 \includegraphics[width=2.1in,angle=0]{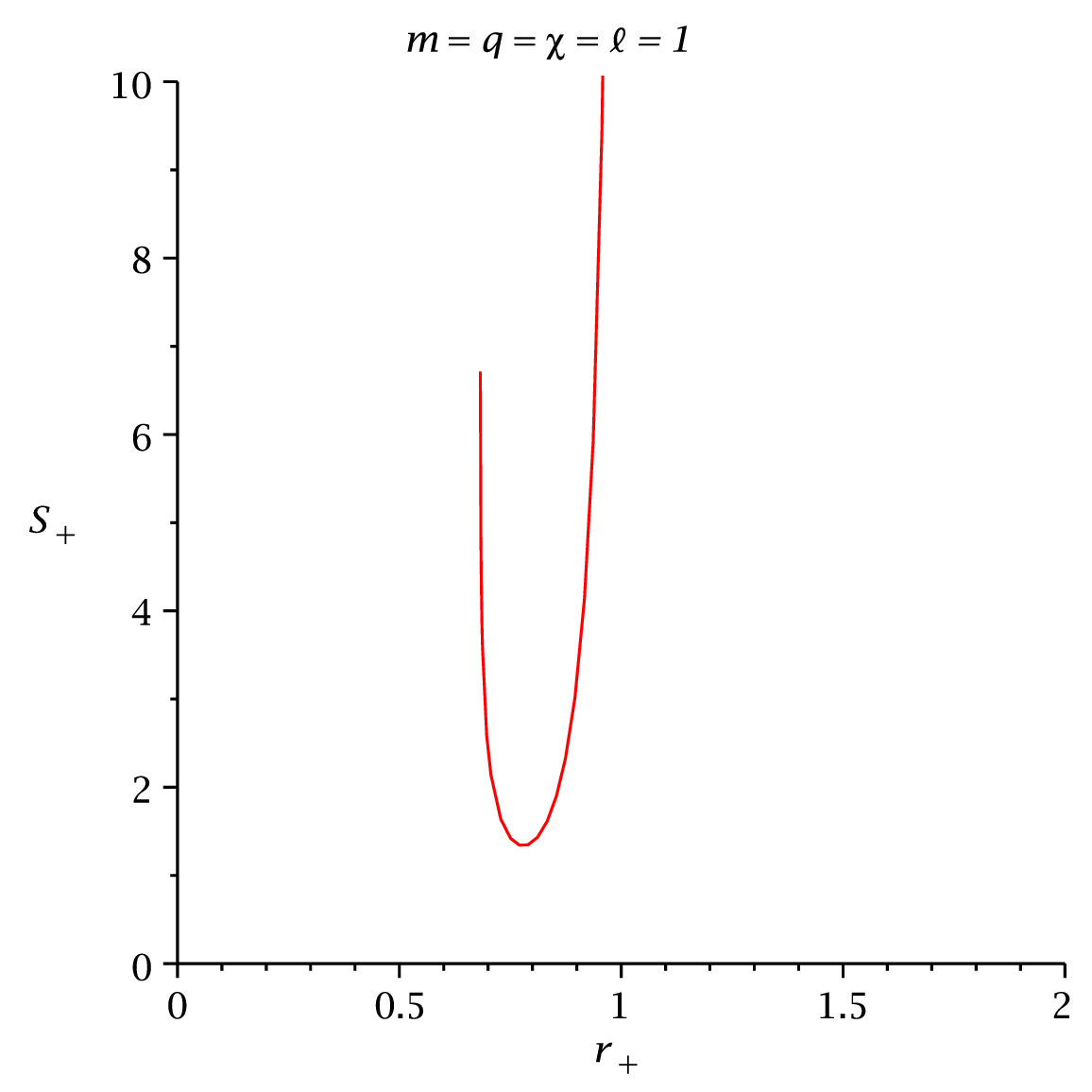}}
 \caption{\label{as3}\textit{ In this graph, we have drawn the variation of logarithmic 
 corrected entropy~(${\cal S}_{+}$)~(right figure) and without 
 logarithm corrected entropy~(${\cal S}_{0, +}$)~(left figure) with horizon radius~($r_{+}$).}}
\end{center}
\end{figure}

We have plotted the logarithmic corrected entropy and without logarithmic corrected entropy in Fig.~\ref{as3}. 
It follows from the graph~[Fig. 3(a)~(left panel)] when there is no acceleration i. e. $\chi=0$ and there is 
no logarithmic correction, the entropy is increasing when horizon radius increases. When there is an 
acceleration the value of entropy diverges at a certain horizon radius this indicates the phase 
transition occurs at $r_{+}=1$.  On the other hand, when we have taken 
the logarithmic correction of entropy in the presence of acceleration then 
the divergence disappears [Fig. 3(b)(right panel)] 
and all the phenomena occurs at $0.5<r_{+}<1$. Three situations are also 
qualitatively different. This is an another \emph{interesting feature} of 
charged accelerating BH.  

\section{\label{dis} Conclusion:}
In this work, we studied the thermodynamic properties of slowly accelerating BH consists of five parameters, namely 
the mass, the charge, the acceleration, the cosmological constant and cosmic string tension. We derived the 
geometric product formula i.e. area  product formula of OH and IH for charged accelerating BH. We showed that 
this area product formula should \emph{not be mass-independent} nor \emph{does it quantized}. This suggests 
that the mass-independent conjecture of Ansorg-Hennig \emph{breaks down} for charged accelerating BH. 
This is an another example we have added in the literature that the area product of two physical 
horizons is \emph{not} always mass-independent. We also derived the famous Cosmic Censorship Inequality 
for this slowly accelerating BH. The physical significance of this inequality is to determine the lower
bound of mass or energy for a time-symmetric initial data which fulfilled the dominant energy condition. 

Moreover, we evaluated the criterion under which the BH showed the second order phase transition.  
Finally, we computed the logarithmic correction to entropy due to  statistical quantum  thermal 
fluctuations near the BH equilibrium temperature for this charged accelerating BH. The implication 
of the logarithmic corrections to the BH entropy may useful to understanding the Suskind's holographic 
hypothesis~\cite{leo} and the AdS/CFT correspondence which is a prime example of holography. This principle 
is based on string theory and quantum gravity. It would be an interesting if one could compute the 
quasilocal energy for this BH following the work~\cite{vir99}. It might be helpful to testify
the Seifert conjecture for BH naked singularity~\cite{vir99}.

\section*{Acknowledgements}
I  would like to thank Prof. Ruth Gregory for useful correspondence. I also would like to 
thank Dr. Marco Astorino for useful comments.

\end{document}